\RequirePackage{fix-cm}
\documentclass[smallextended]{svjour3}       
\smartqed  
\usepackage{graphicx}
\usepackage{epstopdf}
\usepackage{sidecap}
\usepackage{amsmath}
\newcommand{\req}[1]		{(\ref{#1})}

\newcommand{\p}			{\partial}
\newcommand{\x}			{\vec{x}}
\newcommand{\CA}		{\mathcal{A}}
\newcommand{\CL}		{\mathcal{L}}

\newcommand{\measure}		{\!dt\,d^{2\!}x\,N\!\sqrt{g}\,}
\newcommand{\hatmeasure}	{\!dt\,d^{2\!}x\,\hat N\!\sqrt{\hat g}\,}

\begin{document}

\title{GR 20 Parallel Session A3: Modified Gravity}


\author{Petr Ho\v{r}ava     \and   
   Arif Mohd \and Charles~M.~Melby--Thompson \and Peter~Shawhan     
}

\authorrunning{Ho\v{r}ava, Mohd, Melby--Thompson, Shawhan } 

\institute{Petr Ho\v{r}ava \at
              Berkeley Center for Theoretical Physics,
              Department of Physics, University of California,
     Berkeley, CA 94720-7300, USA; 
and 
Physics Division, Lawrence Berkeley National Laboratory,
     Berkeley, CA 94720, USA,
              \email{horava@berkeley.edu}           
\and        
    Arif Mohd  \at
              SISSA - Scuola Internazionale Superiore di Studi Avanzati
                                     and
     INFN sezione di Trieste, Trieste, Italy.
              \email{arif.mohd@sissa.it}       
\and Charles M. Melby--Thompson \at 
Kavli Institute for the Physics and Mathematics of the Universe (WPI),
              The University of Tokyo, Kashiwanoha, Kashiwa, 277-8583, Japan
              \email{charles.melby@ipmu.jp}
\and Peter Shawhan \at University of Maryland, College Park, MD 20742, USA \email{pshawhan@umd.edu}
}

\date{Received: date / Accepted: date}

\maketitle

\begin{abstract}{
The parallel session (A3), on ``Modified Gravity'', enjoyed one on the
largest number of abstract submissions (over 80), resulting in the selection
of 24 oral presentations.  The three short papers presented in the following sections
 are based on the session talks by Arif Mohd on Thermodynamics of universal horizons in Einstein-\AE ther theory,  Conformal anomalies in Ho\v{r}ava-Lifshitz gravity by  Charles Melby-Thompson
and Detectability of scalar gravitational waves by LIGO and Virgo by Peter Shawhan.
They have been selected as a representative sample,
to illustrate some of the best in the remarkable and encouraging variety of topics discussed in the
session -- ranging from highly theoretical, to phenomenological,
observational, and experimental -- with all these areas playing an integral
part in our quest to understand the limits of standard general relativity.
}
\end{abstract}


\bigskip

\section{Einstein-{\AE}ther theory: 
Thermodynamics of Universal horizons - \\
{\small{~~~ Arif Mohd }}} 




{The Noether charge method {\it \`a la} Wald is used  to show that a first law, which resembles the first law of thermodynamics, can be formulated for universal horizons in asymptotically flat, static, spherically-symmetric solutions of Einstein-{\AE}ther theory. }


In the theories of gravity that violate the local Lorentz invariance, like Einstein-{\AE}ther theory and Ho\v{r}ava-Lifshitz gravity, conventional Killing horizon  does not capture the notion of a black hole. This is so because all particles couple to the preferred frame and acquire nonrelativistic superluminal dispersion relations that allow them to penetrate the Killing horizon from inside and escape to infinity. However, it has recently been found \cite{Blas:2011ni} that the asymptotically flat, static, spherically-symmetric solutions of these theories have  a special spacelike hypersurface which traps the modes of arbitrarily high velocities. Signals from beyond this hypersurface can never escape to infinity and are destined to hit the singularity. Hence, this hypersurface is the true causal boundary of spacetime and is called the Universal Horizon. \par
We use the Noether charge method {\it \`a la} Wald to show that a first law, which resembles the first law of thermodynamics, can be formulated for universal horizons in the Einstein-{\AE}ther theory. Given that the attempt to prove a first law for the Killing horizon has been unsuccessful \cite{Foster:2005fr} owing to the irregularity of bifurcation surface ({\ae}ther diverges there), we suggest that in Lorentz violating theories one should ascribe the thermodynamic properties to the universal horizon and not to the Killing horizon. This would also cure the violations of the Generalized Second Law in these theories, which crucially depend upon ascribing the thermodynamic properties to the Killing horizon. 
\subsection{Einstein-{\AE}ther theory}
\label{sec:EA theory}

Einstein-{\AE}ther theory is a generally covariant theory of gravity that violates the local Lorentz invariance due to the presence of a preferred vector called the {\ae}ther, $u^a$. The {\ae}ther is dynamical but constrained to be unit-timelike. \par
Action for the Einstein-{\AE}ther theory is given by
\begin{equation}
S= \frac{1}{16 \pi G_{\ae}}\int \mathrm{d}^4x  \sqrt{-g}\, \left( R + L_{\ae} \right),
\end{equation}
where the {\ae}ther-dependent part is 
\begin{equation}
L_{\ae} = - {Z^{ab}}_{cd} \, \nabla_a u^c \, \nabla_bu^d + \lambda (u^2 +1).
\end{equation}
Here $\lambda$ is  a Lagrangian multiplier that enforces the unit timelike normalization of the {\ae}ther four-vector $u^a$, and ${Z^{ab}}_{cd}$ describes the coupling of the {\ae}ther with the metric in terms of the coupling constants $c_i, i=1,2,3,4$, as
\begin{equation}
{Z^{ab}}_{cd} = c_1\, g^{ab} g_{cd} + c_2\, \delta^a_c \delta^b_d + c_3 \,\delta^a_d  \delta^b_c - c_4\, u^a u^b g_{cd}.
\end{equation} 
The weak-field limit \cite{Carroll:2004ai} can be used to relate constant $G_{\ae}$ occurring in the action and  the Newton's constant G as
\begin{equation}
G_{\ae}= \left( 1 - \frac{c_{14}}{2} \right) G.
\end{equation}
\subsection{Universal horizons}
\label{sec:uh}

\begin{figure}
 \includegraphics[scale=0.15]{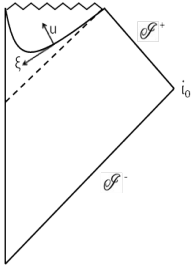}
 \caption{Universal horizon is the hypersurface where the Killing vector field $\xi^a$ becomes orthogonal to the {\ae}ther $u^a$,  and hence is tangential to the hypersurface. } 
\label{fig:uh}
\end{figure}

Consider a static, spherically-symmetric spacetime in which {\ae}ther is hypersurface orthogonal \cite{Eling:2006ec}. At infinity, the {\ae}ther and the time-translation Killing vector $\xi$ are aligned. Inside the Killing horizon $\xi$ becomes spacelike. Consider that particular hypersurface where  $\xi$ becomes orthogonal to the {\ae}ther and hence is tangent to this hypersurface, which is normal to the {\ae}ther (see fig.~\ref{fig:uh}). Any causal signal (i.e., one which propagates in the future of this hypersurface) necessarily moves towards a decreasing radius, and eventually hits the singularity. This hypersurface thus  acts like a one-way membrane, and is called the Universal horizon. Since it traps the modes of arbitrarily high velocities, the universal horizon defines a causal boundary and hence the black hole region in  the spacetime. A regular universal horizon is found to exist in the one-parameter family of asymptotically flat, static, spherically-symmetric solutions of the Einstein-{\AE}ther theory \cite{Eling:2006ec,Barausse:2011pu}. Universal horizons in Ho\v{r}ava gravity are discussed in ref.~\cite{Blas:2011ni}. \par
\subsection{First-law for universal horizons}

Walds' method of proving the first-law for Killing horizon requires that all the fields in the theory are Lie-dragged by the Killing vector field $\xi^a$ and that all the fields admit a regular extension to the bifurcation surface. Since the Lie-drag acts like a radial boost on the bifurcation surface the only vectors that can be Lie-dragged are the ones tangent to it and hence are spacelike. But the {\ae}ther is constrained to be timelike, hence it cannot be Lie-dragged. If we impose that it be Lie-dragged then it must diverge. Hence the bifurcation surface is not regular and one could not prove the first law for Killing horizon in the Einstein-{\AE}ther theory \cite{Foster:2005fr}. We can, however, prove the first law for universal horizon. \par
Recall that in the diffeomorphism invariant theories, in the canonical framework, if the Hamiltonian corresponding to the evolution along vector field $\xi$ exists on the phase space then its on-shell variation is a surface integral,
\begin{equation}
\label{eq:hamiltonian_wald}
\delta H_\xi = \int_{\partial \Sigma} \mathrm{d} \sigma_{ab}\, \left( \delta Q_\xi^{ab} - \,\theta^{[a}\xi^{b]}\right),
\end{equation}
where $Q_\xi^{ab}$ is the Noether charge density and $\theta^a$ is the symplectic potential density whose explicit expressions can be obtained from the variations of the action functional. Let us now take $\xi$ to be the Killing vector field which is  asymptotic time translation and is orthogonal to the {\ae}ther on the universal horizon, and let $\Sigma$ be a slice which cuts the universal horizon orthogonal to $\xi$. Then contribution from spatial infinity is the change in energy and eq.~\eqref{eq:hamiltonian_wald} corresponds to the first-law for the universal horizon. 
We require that on the universal horizon $\delta u^a = 0$ and $\delta g_{ab} = \delta \gamma_{ab}$, where $\gamma_{ab}$ is the induced metric on the cross-section of the universal horizon orthogonal to the spatial vector $s^a$, i.e., $\gamma_{ab} = g_{ab} + u_a u_b - s_a s_b$. In particular, $\delta s^a$ is also equal to zero on the universal horizon.
Restricting the solutions and the perturbations to  one-parameter family of asymptotically flat, static, spherically-symmetric spacetimes we get,
\begin{equation}
\label{eq:delta hamiltonian_UH}
\delta \mathcal{E} = \frac{1}{8 \pi G_{\ae}}\left[\kappa_{UH}(1-c_{13}) + \frac{c_{123}}{2} K_{UH} \|\xi\|_{UH} \right] \delta (\oint \mathrm{d}^2x \sqrt{\gamma}).
\end{equation}
Here, $\kappa_{UH}=\sqrt{- \frac{1}{2}(\nabla_a \xi_b)(\nabla^a \xi^b)}$ is  ``surface gravity" and $K_{UH}$ is the extrinsic curvature of the universal horizon, and $\gamma$ is determinant of the induced metric on the cut of the universal horizon. \par
In ref.~\cite{Berglund:2012bu} one-parameter family of solutions corresponding to two different combinations of the coefficients ($c_{123}=0$ and $c_{12}=0$)  were found. For both the families, the second term in the square brackets in eq.~\eqref{eq:delta hamiltonian_UH} does not contribute. In ref.~\cite{Berglund:2012fk} a tunneling calculation showed that the universal horizon radiates thermally at the temperature given by 
\begin{equation}
T = \frac{\kappa_{UH}}{4 \pi} \sqrt{\alpha \frac{(1-c_{13})}{(2-c_{14})}},
\end{equation}
where, $\alpha$ is $2$ for $c_{123}=0$ and $3$ for  $c_{12}=0$ solution. This suggests that the entropy of the universal horizon of the corresponding solution is given by
\begin{equation}
S_{UH}= \sqrt{\frac{16(1-c_{13})}{\alpha(2-c_{14})}} \,\, \frac{\mathrm{Area}}{4G_N}.
\end{equation}
\subsection{Summary and Outlook}
\label{sec:outlook}

We have used the Noether charge method of Wald to prove the first-law for  universal horizons in 
one-parameter family of asymptotically flat, static, spherically-symmetric black-hole solutions of Einstein-{\AE}ther
theory. It would be interesting to see if one can prove the first law for arbitrary perturbations of the symmetric solutions, just like in general relativity. It would  also be interesting to  have an actual quatum field theory calculation to see if it is indeed the case that  universal horizons radiate thermally at the temperature that the first law suggests.\par
If  it turns out to be the case that the black-hole thermodynamics makes sense only for the universal horizons and not for the Killing horizons in Lorentiz violating theories of gravity, then one would be able to evade all constructions that lead to the violations of GSL in these theories. This subject therefore deserves further study.

\section{The Curious Case of Conformal Anomalies in Ho\v{r}ava-Lifshitz \\
~~~ Gravity -   
{\small{Charles M. Melby--Thompson}}}
The role of anisotropic Weyl anomalies in conformal Ho\v{r}ava-Lifshitz (HL) gravity is considered in 2+1 dimensions.
While classically the scalar mode of HL gravity disappears at the conformal point, experience from the relativistic case suggests that anomalies will re-introduce it at the quantum level.
In this note it is  shown that in HL gravity, the scalar mode can be eliminated even in the presence of a conformal anomaly.
Moreover, when the anomaly takes on a special form it can force gravity into a static phase where the spatial metric undergoes no deformations at all.

\subsection{Conformal Ho\v{r}ava-Lifshitz Gravity}
\label{intro2}

Several years ago, Ho\v{r}ava proposed~\cite{mqc,qglp} a perturbatively unitary, power-counting renormalizable field theory of quantum gravity.
Its improved behavior in the ultraviolet over general relativity is made possible by the addition of higher derivatives to the action; to maintain manifest unitarity, the action is kept quadratic in time derivatives.
This can be done only by giving up local Lorentz invariance.

HL gravity is characterized by a fundamental foliation of spacetime by spatial slices.
In an adapted coordinate system $(t,\x)$, the allowed coordinate transformations take the form
$t\mapsto f(t)$, $\vec x \mapsto \vec g(t,\vec x)$.
The metric decomposes under these transformations as
\begin{equation}
ds^2 = -N^2 dt^2 + g_{ij}(dx^i+N^idt)(dx^j+N^jdt) .
\end{equation}
With the loss of a local symmetry comes an extra propagating scalar degree of freedom.
I restrict myself here to 2+1 dimensions, where it is the only graviton polarization.

The symmetries of the theory also permit an extended group of local transformations, the anisotropic Weyl transformations~\cite{mqc}
\begin{equation}
\label{eq:awt}
\tilde N = e^{z\omega} N
\qquad
\tilde N_i = e^{2\omega} N_i
\qquad
\tilde g_{ij} = e^{2\omega} g_{ij} \,.
\end{equation}
I will use the word ``conformal'' in this extended sense: conformal HL gravity~\cite{mqc,qglp} (with exponent $z$) is thus any theory invariant under~\req{eq:awt}.
The most general conformally invariant action in 2+1 dimensions that is (1)~quadratic in time derivatives and (2)~consistent with the symmetries has $z=2$, and is given by:
\begin{equation}
S_\mathrm{CHL}[N,N_i,g]=\int\measure(|K|^2-\kappa \tilde R^2) ,
\end{equation}
with $|K|^2=K^{ij}K_{ij}-\frac12K^2$, $K_{ij}=\frac{1}{2N}(\dot g_{ij}-\nabla_iN_j-\nabla_jN_i)$ the extrinsic curvature of spatial slices, and $\tilde R=R-\frac{(\nabla N)^2}{N^2}+\frac{\nabla^2N}{N^2}$.

Classically, this theory has no propagating degrees of freedom as a result of Weyl invariance, but even in 2+1 dimensions anisotropic Weyl invariance is often spoiled by anomalies at the quantum level~\cite{nr-weyl,lgh,baggio}.
Understanding the conformal anomaly and its gravitational implications is crucial to formulating quantum conformal HL gravity, and will occupy the remainder of this paper.

\subsection{Ho\v{r}ava-Lifshitz Gravity and the Conformal Anomaly}

Consider the path integral for conformal HL gravity.
By first integrating out all but the metric fields, it can be written
\begin{equation}
Z_\mathrm{CHL}=\int\! dN\,dN_i\,dg_{ij} \, e^{iS_\mathrm{CHL}[N,N_i,g_{ij}]} Z_\mathrm{m+gf}[N,N_i,g_{ij}]
\end{equation}
where $Z_\mathrm{m+gf}$ is the partition function for the matter and gauge-fixing/ghost sectors.
(From here on, we will refer to all non-metric fields as ``matter''.)

To maintain conformal invariance after coupling to matter, the matter sector must also be conformally invariant.
Classically this means the action is invariant under conformal rescalings, but quantization spoils this invariance due to the anomalous variation of the path integral measure.
Because anomalies arise from the regularization of ultraviolet divergences, the anomalous transformation of the matter sector path integral is given by a local expression:
\begin{equation}
Z[e^{2\omega}N,e^{2\omega}N_i,e^{2\omega}g_{ij}]=Z[N,N_i,g_{ij}]e^{i\int\measure W(\omega;N,N_i,g_{ij})} ,
\end{equation}
with $W$ a {\em local} function of its arguments and their derivatives.
The anomaly is determined uniquely by its infinitesimal form
\begin{equation}
\delta_\omega Z[N,N_i,g_{ij}] = i \int\measure \omega \CA(N,N_i,g_{ij}) \, Z[N,N_i,g_{ij}] .
\end{equation}
Terms appearing in the infinitesimal anomaly must satisfy two conditions:
(1)~Scale invariance, and (2)~the Wess-Zumino consistency condition ({\it i.e.} the anomaly is integrable).
These conditions were applied in~\cite{lgh} to classify the most general anomaly, which is a linear combination of 8 terms.

Some anomaly terms can be eliminated by adding gravitational terms to the matter action.
These have no effect on the matter dynamics, but modify the gravitational action.
The anomaly is thus only defined uniquely modulo gravitational counterterms.
A basis for the anomaly is given by the two quadratic terms
\begin{equation}
\label{irreducible-anomaly}
|K|^2
\qquad
\tilde R^2
\end{equation}
Both anomalies can in fact be realized using a free scalar field: the conformal Lagrangian
$\CL = N^{-2}[(\p_t-N^i\nabla_i]\phi)^2-(\Delta\phi)^2-\alpha\tilde R^2\phi^2$
gives rise to the anomaly $\CA = \frac{1}{32\pi}|K|^2-\frac{\alpha}{8\pi}\tilde R^2$~\cite{baggio,lglh}.



\subsubsection*{Conformal anomaly as constraint}
\medskip
As is always the case, we can obtain a theory that is trivially conformal by introducing an auxiliary field $\sigma$ and defining new metric variables by $(N,N_i,g_{ij})=e^{2\sigma}(\hat N,\hat N_i,\hat g_{ij})$. 
The path integral is now invariant under conformal transformations in which $\delta(\hat N,\hat N_i,\hat g_{ij})=2\omega(\hat N,\hat N_i,\hat g_{ij})$ and $\delta\sigma=-\omega$.
After gauge fixing this conformal symmetry by some convenient condition (such as
$\det\hat g=1$),
the dependence of the path integral on the scale factor is now captured by the path integral over $\sigma$.

The anisotropic Weyl anomaly $\CA$ in 2+1 dimensions has a crucial property: it is itself Weyl invariant.
In this case, the finite form of the partition function variation $i\int\measure \delta\omega \CA$ is simply the phase factor $\exp\left(i\int\measure\omega\CA\right)$.
Using the integrated form of the anomaly, we see that the effect of the path integral over $\sigma$ is to impose a constraint:
\begin{equation}
\int D\sigma\,e^{i\int\hatmeasure\sigma\,\CA(\hat N,\hat N_i,\hat g_{ij})}\propto\delta(\CA) .
\end{equation}

Consider now a situation in which all but the $|\hat K|^2$ anomaly term vanishing, giving the constraint $|\hat K|^2\equiv 0$.
(This is possible, since all but the two terms of~\req{irreducible-anomaly} can be eliminated by choice of action, and the $\tilde{R}^2$ anomaly can be set arbitrarily within the free field action.)
The constraint can be put in a more enlightening form by a particular choice of gauge.
At any point $(t_0,\x_0)$, choose a gauge such that 
\begin{equation}
\hat N(t_0,\x_0)=1
\quad
\hat N_i(t_0,\x_0)=0
\quad
\hat g_{ij}(t_0,\x_0)=\delta_{ij} .
\end{equation}
We also choose a conformal frame in which $g^{ij}\dot g_{ij}=0$, {\it i.e.} $\dot g_{xx}=-\dot g_{yy}$.
Then at $(t_0,\x_0)$ the constraint takes on the form 
\begin{equation}
|\hat K|^2 = \dot{\hat g}_{xx}^2+2\dot{\hat g}_{xy}^2 = 0 .
\end{equation}
Using the relation $\delta(x^2+y^2)=\frac{\pi}{2}\delta(x)\delta(y)$, we see that up to normalization, $\delta(|\hat K|^2)=\delta(\dot{\hat g}_{xx})\delta(\dot{\hat g}_{xy})$ at $(t_0,\x_0)$.
While this is a gauge-fixed condition at one point, its covariantization $\delta^{(2)}(K_{ij}-\frac{1}{2}Kg_{ij})$ holds everywhere.%
\footnote{In interpreting this one must recall that $K_{ij}-\frac{1}{2}Kg_{ij}$ is traceless and thus has two components.}
The anomaly has thus implemented a constraint on the geometry at the level of the path integral.

Let us understand better what the resulting theory looks like.
Use diffeomorphisms and conformal transformations respectively to fix $N_i=0$ and $g^{ij}\dot g_{ij}=0$.
This leaves a residual {\em global} spatial diffeomorphism symmetry, which we use to fix $g_{ij}=\delta_{ij}$ at $t=0$. 
The constraint condition, combined with the choice of conformal gauge, implies $\dot g_{ij}=0$, and so we have $g_{ij}=\delta_{ij}$ everywhere.

All that is left of the metric is $\hat N$.
Because it has no kinetic term, in principle it can be integrated out, leaving a potential under which the matter sector interacts.
This reduces conformal HL gravity with $\CA\propto|K|^2$ to a non-dynamical scalar potential sourced by the kinetic energy of the matter sector, supplemented by a global Hamiltonian constraint.

\subsection{Conclusions}

We have seen that in 2+1 dimensional HL gravity, conformal anomalies have an unexpected effect: rather than resurrecting the propagating scalar mode as happens in relativistic quantum gravity in two dimensions, it implements a constraint at the level of the path integral measure.
In some cases, this constraint is powerful enough to force the spatial metric into a static phase.
These results beg a further work, such as the analysis of the general anomaly and a careful BRST analysis.

Perhaps the most interesting, however, is the case of pure gravity: since the theory has no propagating degrees of freedom, the theory may even be exactly solvable.
These questions will be addressed in part in forthcoming work.

\section{Detectability of scalar gravitational-wave bursts with LIGO 
and  Virgo}
\vskip-.2cm
{\small{{\bf -- P. Shawhan},\footnote{University of Maryland, 
College Park, MD 20742, USA}
{in collaboration with
 S. Sullivan$^2,$\footnote{Present address: University of Colorado, Boulder, CO 80309, USA},
 M. Avara$^2$,
           G. Vedovato\footnote{INFN Sezione di Padova, I-35131 Padova, Italy},
           S. Coughlin\footnote{Northwestern University, Evanston, IL 60208, USA}, M. Drago\footnote{INFN, Gruppo Collegato di Trento and Universit \`{a} di Trento, I-38050 Povo, Trento, Italy}, K. Hayama\footnote{National Astronomical Observatory of Japan, Tokyo 181-8588, Japan},
           I. Kamaretsos\footnote{Cardi ff University, Cardi ff, CF24 3AA, UK},
P. Sutton$^7$, S. Klimenko\footnote{University of Florida, Gainesville, FL 32611, USA}}}}


 In many alternative theories of gravity, gravitational waves (GWs) can
propagate with a scalar polarization mode (and possibly others) in
addition to the two tensor modes predicted by general relativity. The
scalar mode could even be dominant, particularly in the case of
spherical collapse events of stellar cores or neutron stars. No
explicit search for such signals has been carried out yet. In this
work,  the detectability of simulated scalar GW burst
signals by the LIGO-Virgo network is studied using a slightly modified version of
a standard GW burst search pipeline. It is  found that typical scalar burst
signals can be detected well by these pipelines, which motivates
carrying out actual searches in the future. Interestingly, scalar GW
signals can be detected with nearly the same efficiency by an
un-modified (tensor mode) all-sky burst search pipeline, although the
source positions are misreconstructed.


Data from the LIGO and Virgo gravitational-wave (GW) detectors has
been analyzed to search for many types of signals, including arbitrary
GW ``bursts'' \cite{S6VSR2burst}.  All such searches to date have
implicitly assumed that general relativity (GR) is the correct theory
of gravity.  For instance, the standard GW burst search pipelines
``coherent WaveBurst'' \cite{CWB} and ``X-Pipeline'' \cite{XPipeline}
attempt to interpret the GW detector data as a linear combination of
the ``plus'' and ``cross'' tensor polarization modes that GR allows,
plus detector noise.  Relative arrival times and antenna response
factors are computed for the $+$ and $\times$ components, depending on
the direction of arrival.  X-Pipeline is designed to search a known
sky position, such as the location of a gamma-ray burst \cite{S6GRB},
while coherent WaveBurst can efficiently do an ``all-sky'' search
considering all possible arrival directions.

Many alternative theories of gravity allow one or more extra
polarization modes in addition to the two tensor modes \cite{WillLR}.
For instance, a transverse scalar (``breathing'') mode is common in
tensor-scalar theories (including the Brans-Dicke theory as a specific
case) and others.  The ``trace'' mode in Einstein-{\AE}ther theory is a
particular combination of transverse scalar and longitudinal scalar
basis modes (T.\ Jacobson, private communication).  Vector modes are
also allowed by some theories, but will not be discussed here.

If GR is not the correct theory of gravity, the standard analysis of
LIGO and Virgo data could conceivably miss a GW signal by considering only
the two tensor modes.
In this work, we demonstrate that a modified analysis can successfully
recover such signals, and compare to the standard analysis.

\subsection{Detectability study}
\label{study}

The response of an interferometric GW detector to a scalar
polarization mode is well understood \cite{MaggioreNicolis}.  In
principle, \emph{arbitrary} $+$, $\times$ and scalar components could
be disentangled from a GW signal arriving at the Earth, but that would
require more than three detectors and depends on the relative
amplitudes of the components.  For now, we choose to focus on
\emph{just} the scalar mode, supposing that it either dominates the
tensor modes in the received signal, or else arrives at a different
time due to a difference in speeds \cite{WillLR}.  This choice is
motivated by spherical (or nearly spherical) collapse events, such as
a stellar core collapsing to a neutron star (NS) or black hole (BH),
or a NS collapsing to a BH.  Such cases have strong geometric coupling
to the scalar mode; the coupling factor in the theory then will
determine what wave amplitude propagates away.

Some modeling has been done of source dynamics and GW emission in a
general class of tensor-scalar theories allowing nonlinear coupling - for the theoretical background, see for instance
\cite{DEF}.  Scenarios include stellar core collapse \cite{NovakSC},
NS collapse to a BH \cite {NovakNS}, and ``spontaneous scalarization''
of a NS \cite{NovakSS}.  We use a Monte Carlo technique to study
detectability of these modeled signals, as well as ad-hoc
simulated signals such as ``sine-gaussians'' and ``white-noise
bursts'' like those described in \cite{S6VSR2burst} but with scalar
rather than tensor polarization.  We generated signals with random
times and arrival directions and added them to real detector noise
from the LIGO S5 science run, but with the data from the 2-km Hanford
detector ``relocated'' to Virgo.  We analyzed the resulting data
streams with a modified version of the coherent WaveBurst all-sky
search pipeline which considered only a single, scalar polarization
mode with the correct antenna response.

We find that the modified search pipeline successfully detects both
ad-hoc and spherical-collapse waveforms.  Some preliminary results are shown 
in Fig.\ 2.
\begin{figure}
\includegraphics[width=1\textwidth]{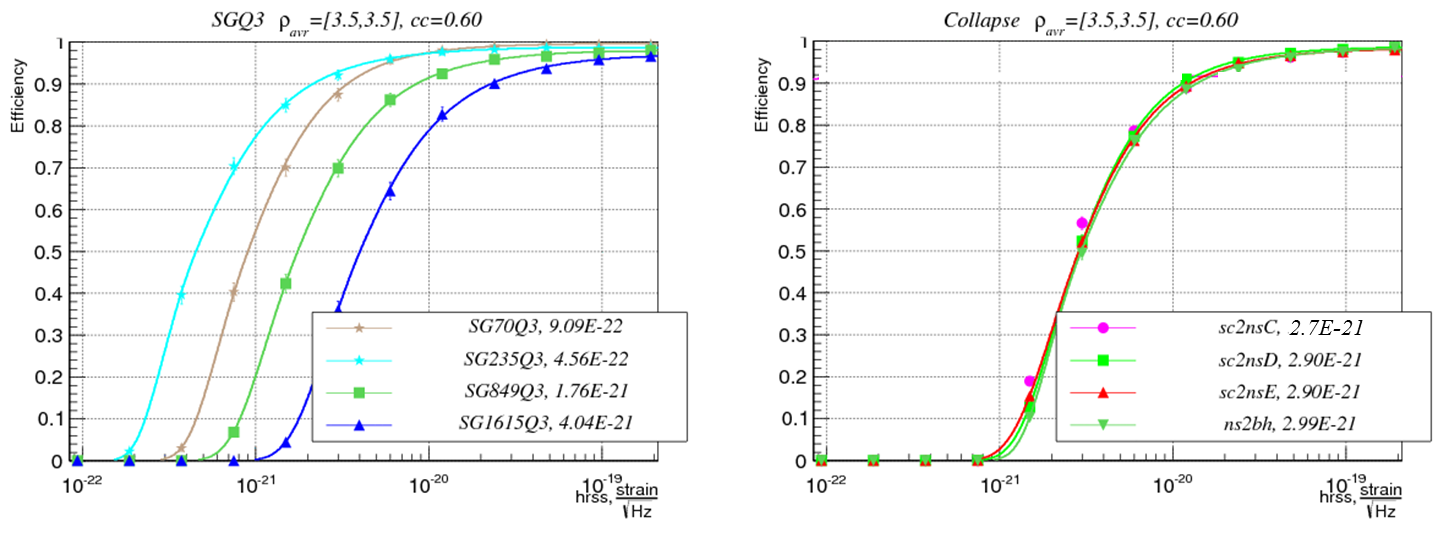}
\caption{Detection efficiency as a function of signal amplitude
  arriving at Earth for ad-hoc ``sine-gaussian'' signals (left) and
  model waveforms representing stellar core and neutron star collapse
  in tensor-scalar theories (right)}
\label{fig:1}       
\end{figure}
The variation in the detection-efficiency curves for sine-gaussians is
consistent with the frequency dependence of the detector
sensitivities.  On the other hand, the nearly equal sensitivities for
the collapse signals reflect the fact that their GW strain signatures
are all more-or-less step functions, with equal spectral content in
the sensitive band of the detectors.
The ``background'' from randomly coincident signal fluctuations seems
to be about the same for this search as for the previously published
all-sky burst search \cite{S6VSR2burst}.


We also ran the standard (tensor) all-sky burst search on the same
(scalar) simulated GW signals, and it turns out that the standard
search can detect these signals nearly as well as the scalar search!
Examining individual events, the reason is clear: the tensor analysis
has the freedom to infer a source position at a different place in the
sky, where the calculated (tensor) antenna responses and relative
timing are evidently similar enough to the actual (scalar) values.
The scalar search is about 30\% more sensitive for the scalar collapse
waveforms, and also somewhat more sensitive for most of the ad-hoc
scalar waveforms.  However, scalar white-noise burst signals are
detected \emph{less} well by the scalar search than by the tensor
search; the reason for this is unclear.

\subsection{Discussion}
\label{discussion}

Existing constraints on the parameter space of nonlinear tensor-scalar
theories from the Cassini timing experiment and binary pulsars
\cite{Freire} limit the possible strength of scalar GWs from spherical
collapse events, essentially ruling out the regime in which
``spontaneous scalarization'' effects would amplify the signal.
Consequently, we estimate that a scalar signal from a stellar-mass
collapse event could only be detected by current GW detectors if it
occurred within our galaxy.

Our detectability study implies that published LIGO-Virgo burst
searches such as \cite{S6VSR2burst} actually place limits on scalar GW
bursts as well as the tensor bursts that they were designed to target.
Some sensitivity could be gained from a separate scalar burst search,
but this has not been undertaken so far.  Future searches can be run
with this option.  We should also work out a statistical framework to
determine whether any given GW burst signal we detect in the future
contains a scalar component or not.

In parallel, we have investigated searching for a scalar GW signal
associated with a known nearby core-collapse supernova using a suitably
modified version of X-Pipeline.  In this case the known sky position
determines the relative arrival times and antenna responses, but those
depend on the sidereal time of the initial collapse, which is poorly
determined by the optical light curve.



\begin{acknowledgements}
PH would like to thank the organizers of the conference, in particular 
Bala Iyer and Jerzy Lewandowski, for their superb organizational efforts, and 
for making our parallel session possible.
AM  would like to thank Abhay Ashtekar, Bethan Cropp, Ted Jacobson, Stefano Liberati, Thomas Sotiriou and Matt Visser for discussions.
CM  would like to thank Petr Ho\v{r}ava for useful discussions.
His work was supported by the World Premier International Research Center Initiative (WPI Initiative), MEXT, Japan.
Work by PS et al was supported by U.S.\ National Science Foundation grant
PHY-1068549.  They thank their LIGO Scientific Collaboration and Virgo
Collaboration colleagues and Jerome Novak for useful discussions.
The GW summary has the identifier LIGO-P1300206-v2.
\end{acknowledgements}


\bibliographystyle{spphys}       
\bibliography{Horavaetal-A3}   
\end{document}